\documentclass[conference]{IEEEtran}
\IEEEoverridecommandlockouts
\usepackage{cite}
\usepackage{float}
\usepackage{amsmath,amssymb,amsfonts}
\usepackage{algorithmic}
\usepackage{graphicx}
\usepackage{numprint}
\usepackage[inline]{enumitem}
\usepackage{booktabs}
\npdecimalsign{.}
\usepackage{textcomp}
\usepackage{subcaption}
\usepackage{listings}
\lstdefinestyle{mystyle}{
    language=C++,              
    basicstyle=\small\ttfamily, 
    keywordstyle=\color{blue}, 
    commentstyle=\color{green},
    numbers=left,              
    numberstyle=\tiny,         
    framerule=0.5pt, 
    numbersep=-2pt,             
    breaklines=true,           
    frame=single,              
    tabsize=2                  
}
\lstdefinestyle{CStyleCompactSmall}{
    backgroundcolor=\color{backgroundColour},
    commentstyle=\color{mGreen},
    keywordstyle=\color{magenta},
    numberstyle=\tiny\color{mGray},
    stringstyle=\color{mPurple},
    basicstyle=\scriptsize\ttfamily, 
    breakatwhitespace=false,
    breaklines=true,
    captionpos=b,
    keepspaces=true,
    numbers=left,
    numbersep=2pt, 
    showspaces=false,
    showstringspaces=false,
    showtabs=false,
    tabsize=2, 
    language=C
}
\usepackage[table,xcdraw]{xcolor}
\usepackage[framemethod=tikz]{mdframed}
\usepackage{balance}
\mdfdefinestyle{mpdframe}{
    frametitlebackgroundcolor   =black!15,
    frametitlerule          =true,
        roundcorner     =8pt,
        middlelinewidth     =1pt,
        innermargin     =0.25cm,
        outermargin     =0.25cm,
        innerleftmargin     =0.25cm,
        innerrightmargin        =0.25cm,
        innertopmargin      =2mm,
        innerbottommargin   =2mm,
            }
\mdfdefinestyle{insight}{%
        style=mpdframe,
        frametitle={insight},
    }
\usepackage{adjustbox}
\usepackage{changepage} 
\usepackage{hyperref}
\usepackage{multirow}
\newcommand{\TODO}[1]{\textcolor{blue}{#1}\GenericWarning{}{LaTeX Warning: TODO: #1}}\newcommand\todo\TODO
\usepackage{comment}
\definecolor{dkgreen}{rgb}{0,0.6,0}
\definecolor{gray}{rgb}{0.5,0.5,0.5}
\definecolor{mauve}{rgb}{0.58,0,0.82}
\title{In industrial embedded software, are some compilation errors easier to localize and fix than others?\\
\footnotesize \textsuperscript{}
\thanks{This work was partially supported by the Wallenberg Artificial Intelligence, Autonomous Systems and Software Program (WASP) funded by the Knut and Alice Wallenberg Foundation.}
}

\makeatletter
\newcommand{\linebreakand}{%
  \end{@IEEEauthorhalign}
  \hfill\mbox{}\par
  \mbox{}\hfill\begin{@IEEEauthorhalign}
}
\makeatother

\author
{\IEEEauthorblockN{Han Fu\IEEEauthorrefmark{1}\IEEEauthorrefmark{2},
Sigrid Eldh\IEEEauthorrefmark{1}\IEEEauthorrefmark{3},
Kristian Wiklund\IEEEauthorrefmark{1},
Andreas Ermedahl\IEEEauthorrefmark{1}\IEEEauthorrefmark{2},
Philipp Haller\IEEEauthorrefmark{2} and
Cyrille Artho\IEEEauthorrefmark{2}}
\IEEEauthorblockA{\IEEEauthorrefmark{1}\textit{Ericsson AB,}
Stockholm, Sweden \\
Email: \{han.fu,\,sigrid.eldh,\,kristian.wiklund,\,andreas.ermedahl\}@ericsson.com
\IEEEauthorblockA{\IEEEauthorrefmark{2}
\textit{KTH Royal Institute of Technology,}
Stockholm, Sweden \\
Email: \{phaller,\,artho\}@kth.se
}
\IEEEauthorblockA{\IEEEauthorrefmark{3}
\textit{Mälardalen University,}
Västerås, Sweden\\
}}}

\begin{document}
\maketitle
\begin{abstract}
Industrial embedded systems often require specialized hardware. However, software engineers have access to such domain-specific hardware only at the continuous integration (CI) stage and have to use simulated hardware otherwise. This results in a higher proportion of compilation errors at the CI stage than in other types of systems, warranting a deeper study.

To this end, we create a CI diagnostics solution called ``Shadow Job'' that analyzes our industrial CI system. We collected over 40000 builds from 4 projects from the product source code and categorized the compilation errors into 14 error types, showing that the five most common ones comprise 89\,\% of all compilation errors. Additionally, we analyze the resolution time, size, and distance for each error type, to see if different types of compilation errors are easier to localize or repair than others.

Our results show that the resolution time, size, and distance are independent of each other. Our research also provides insights into the human effort required to fix the most common industrial compilation errors. We also identify the most promising directions for future research on fault localization. 

\end{abstract}

    \begin{IEEEkeywords}
    continuous integration, software build, compilation error, fault localization
    \end{IEEEkeywords}

\section{Introduction}\label{introduction}
Agile development~\cite{abrahamsson2017agile} relies on continuous integration (CI)~\cite{chen2015continuous}. In large industries, many agile teams submit code changes concurrently. A previous paper~\cite{fu2022prevalence} shows that dependency issues in hardware-in-the-loop testing result in a significant portion of build failures. This is because the hardware platform is developed alongside the software system. Toward the end of the development cycle, hardware prototypes are made available to the CI test bed. Software developers can then integrate their changes with the new platform. The tests of that integration can only be done in full on the CI platform, as it would be prohibitively expensive to procure a hardware prototype for each developer.

Therefore, industrial hardware-software co-development fundamentally differs from typical open-source projects often studied in the literature. Compilation errors occur very frequently at this integration stage and are responsible for most of the CI failures~\cite{fu2022prevalence}. 

Our goal is to investigate opportunities to enhance fault localization and program repair by addressing these problems in this unique setting and streamlining the process. To locate a faulty module, analyzing and repairing the error can account for a significant part of the development effort, leading to lower productivity, higher costs, and a potentially longer time to market~\cite{jonsson2013increasing,kerzazi2014factors}. Hence, automation of fault localization can be a key differentiating ability for any business or activity~\cite{wong2016survey}.

In this paper, we set out to investigate the main research question, which has five sub-research questions: \textbf{RQ.}~\textit{What are the key factors contributing to compilation errors in the context of industrial embedded system CI?}\label{RQ}

We studied the localization and resolution of compilation errors in industrial embedded systems to identify the potential of automatic remedies for typical errors. 

Our proposed CI diagnostics solution, called ``Shadow Job'', is designed to address our research questions. We analyze over 40000 builds and identify 14 compilation error types representing 98\,\% of the failed compilations. Amongst all errors, five most common ones comprise 89\,\% of all compilation errors. The main reason for the high rate of compilation errors is the disparity in the hardware and software development setups, prior to the CI compilation process. Our results shows the potential of adopting fault localization and automatic program repair techniques for these kinds of issues in CI systems. 

Our contribution focuses on conducting a thorough analysis of compilation errors to inform our efforts in developing automatic solutions. This analysis encompasses spatial and temporal aspects, including the examination of resolution times, size, and distance by tracing the corresponding fixes. 

These metrics provide information about the time taken to resolve errors, the extent of code modifications required, and the spatial distribution of fixes within the code base. Our analysis reveals that the long resolution times for frequent error types do not necessarily result in larger resolution sizes. 

Our paper is structured as follows: Section~\ref{sec:background} presents the background and motivation of our research. Section~\ref{sec:related-work} provides an overview of related work on compilation errors in CI and dependency errors. Section~\ref{sec:studydesign} details our study design, outlining the employed methodologies. Section~\ref{sec:study-results} presents the results of our study. In Section~\ref{sec:conclusion}, we draw conclusions based on our findings. Section~\ref{sec:futurework} outlines future work.

\section{Background}\label{sec:background}
Continuous Integration (CI) automates the integration of code changes from multiple contributors and teams into a shared software project. CI revolves around regularly merging code changes into a central repository, triggering builds and tests through automated tools.
\subsection{Industrial context}
For large industrial embedded systems, it is common for teams to submit a high volume of code to serve multiple product variations in a centralized CI system. Despite successful local compilation and testing, developers may encounter several dependency issues during compilation in the CI. 

\begin{figure}[H]
    \centering  
    \includegraphics[scale=.8]{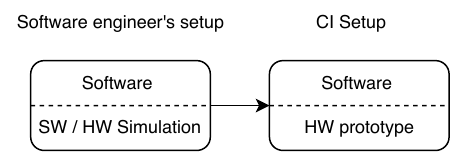}
    \caption{Development Dependency}
    \label{fig:DevDependency}
\end{figure}

Fig.~\ref{fig:DevDependency} illustrates a notable distinction between the setups used by software engineers and the CI environment in a hardware and software co-development context. Most effort is spent on integrating the software with the hardware~\cite{antinyan2014identifying}.

During the software development phase, engineers make use of both software and hardware simulators to establish essential functionality. Subsequently, they transition to submitting their code into the CI loop, where it undergoes quick and thorough testing in conjunction with a hardware prototype. The adoption of a centralized CI environment, serving both software and hardware, offers convenience for globally distributed development teams. Unfortunately, this unified setup often introduces an interface mismatch between the software engineer's initial setup and the CI environment.

Many industry projects involve distributed teams, where team members are working in different locations or time zones. Miscommunications or misunderstandings about code changes can lead to undeclared items when changes are integrated. 

Due to the fast-paced and intense delivery cycle, the development of hardware and software is typically conducted asynchronously. To aid developers, prioritization is given to the setup of hardware in the centralized Continues Integration (CI)  environment. In the early stages, the software team primarily relies on simulators for development, compilation, and local testing. A mismatch becomes evident between hardware and software development, when developers commit to the CI. The discrepancy is characterized by a significant misalignment in development progress or objectives, which can pose challenges for achieving synchronization and integration between hardware and software components. 

The gap between the CI build system and the local development environment cannot be ignored. Addressing this disparity is crucial for ensuring consistent software behavior and efficient debugging in the context of our development process.
In the context of this study, we introduce Listing~\ref{codeExample} as an illustrative case exemplifying a communication breakdown between the hardware and software development. Within this example, the variable \texttt{err} is defined in a separate file. 

\begin{lstlisting}[style=mystyle, caption={Code Example}, label=codeExample]
    Error: cannot convert 'ProductError' to 'const char*'
    Error code: 
        CATCH_THROW_ERROR(err);
    Fixing code: 
        if (err)
        {TRACE_ERROR(SSTR(err)); return false;}
\end{lstlisting}

Notably, the error message in this instance underscores an attempt to pass a \texttt{ProductError} object as an argument to a function explicitly designed to accept a \texttt{const char*}. The \texttt{ProductError} object originates from the hardware design. However, the original software design did not account for the \texttt{ProductError} attribute, resulting in a failure within the CI process. Consequently, this incongruity leads to a failure within the CI compilation process. Our findings shed light on the resource requirements for resolving various compilation errors, providing developers with insights to effectively prioritize their efforts in addressing these issues. 

\subsection{The CI pipelines under study}
Continuous integration seeks to make integrating changes from multiple contributors easier by sharing smaller updates more frequently. The developer submits a code change to be integrated with other system components, which is then tested to ensure that no regression has occurred. It is possible to quickly provide feedback to the developer on the quality of a code change, with the correct architecture and test planning.

Implementing a CI system for embedded software can provide immediate feedback, reducing error correction costs when multiple teams are working on the same project~\cite{planning2002economic}. The CI process in this context is commit-based, with a build triggered as soon as a commit is pushed.

\begin{figure}
    \centering  
    \includegraphics[scale=.6]{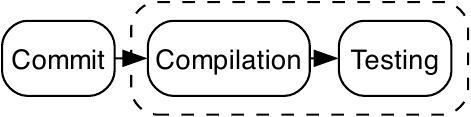}
    \caption{CI pipeline}
    \label{fig:pipeline}
\end{figure}

As shown in Fig.~\ref{fig:pipeline}, a~\emph{pipeline} is a series of actions that run sequentially, and the actions are always run together as an atomic unit. The steps that form the CI pipeline consist of distinct subsets of tasks named commit, compilation, testing, and delivery. The CI under study is running with software that is written in C/C++.  

Our study was centered on four key projects: three test frameworks and one core product, each with a decade-long implementation history. This deliberate selection offers us a holistic perspective on the Ericsson CI system. All four projects undergo testing and verification using the same CI infrastructure, despite involving globally distributed developers and development teams. This setup enables the shadow job machinery to comprehensively collect data across all projects.

\subsubsection*{Definitions} 
A~\emph{commit} is a set of files that pushes the latest changes of the source code to the main branch. A~\emph{patch} is a file containing the set of differences between two versions.\footnote{As the project involves several repositories, we consider a patch to be the union of the outputs of each intra-repository~\textsl{git diff} command.} A~\emph{patch} is associated with a~\emph{commit}. \emph{Compilation} consists of code formatting, static analysis, and the actual compilation. \emph{Testing} executes unit and integration tests on the product. 

\section{Related work}\label{sec:related-work}
CI has seen substantial recent development, capturing attention from both research and industry. Hilton et al.~\cite{hilton2016usage} find that CI is widely used within the most popular open-source projects. Studying its usage, cost, and benefits, they find that CI helps projects release twice as often. Vasilescu et al.~\cite{vasilescu2015quality} study 246 GitHub projects using CI aiming for productivity and quality, showing that CI helps detect more bugs by core developers. Vassallo et al.~\cite{vassallo2017tale} compare CI build errors in 349 Java open-source projects and 418 projects in an industrial organization. They reported that open-source projects' most frequent failure types are testing, compilation, and dependency issues. Furthermore, industrial projects' most frequent failure types are testing, release preparation, and static analysis. 

Beller et al.~\cite{beller2017oops} investigate the impact of CI and conclude that CI build failures are caused by test failures. The test execution part of the test-debug-fix cycle is already automated to a large extent in the software industry~\cite{garousi2017test}. This is partially due to the agile ``revolution'', which had brought expectations of quick feedback to developers on product quality. As a result, continuous integration has emerged as a critical productivity-enhancing tool in the software industry over the last decade~\cite{stolberg2009enabling}.

Relatively little research has been conducted on compilation errors in industrial CI environments. 

\subsection{Compilation errors in CI}
The feedback from the compiler is one of the most important influences in software development. Zhang et al.~\cite{zhang2019large} study compilation errors on~\numprint{3799} open-source projects. They investigate the most common compilation error types and their fix time. They manually analyze 325 broken builds to summarize fix patterns of the ten most common compilation error types. Previous research~\cite{becker2019compiler} has highlighted that compilation error messages can be challenging to interpret and may not be as effective, especially for novice developers. Compilation error messages can be notoriously difficult to comprehend, as shown by Rosen et al.~\cite{rosen1965pufft}. In the context of novice programmers, the work of Traver~\cite{traver2010compiler} shows that compilers can detect common programming errors, but they usually do not pinpoint error locations accurately. 

Seo et al.~\cite{seo2014programmers} studied compilation errors in Google's build process, focusing on Java and C++ environments in a software-centric, cloud-based system. Our research identifies dependency issues as the main cause of compilation errors, constituting 84\,\% of errors across 7 categories. While aligning with Seo's findings, our embedded system and fully automated remote CI system result in significantly different dependency error proportions.

Previous research has shown the prevalence of compilation errors in industrial CI systems~\cite{fu2022prevalence}. Notably, our study investigates a hardware-based platform, unlike Seo's study, which was based on a purely software platform. Our results align with Seo's in terms of error categorization, as we also classify errors into 5 categories. However, our approach to calculating resolution time differs from Seo's methodology. Specifically, we calculate resolution time from the last failure CI build, preceding the successful build, rather than from the first failure CI build. This modification allows us to mitigate the influence of unrelated code changes in our analysis. These differences underscore the importance of our investigation within our specific context.

Barrak et al.~\cite{barrak2021builds} study on \numprint{27675} Travis CI builds of 15 GitHub projects. They identified that features from the build history, author, code complexity, and code/test smell dimensions are the most important predictors of build failures. Ivens et al.\cite{silva2023factors} conducted an analysis of 18 industrial projects within a software company, wherein they calculated 13 metrics for each project based on existing literature related to build failure analysis. Their findings revealed significant correlations between the factors under study and the duration required for correcting build failures. They observed that build failures involving a higher number of modified lines of code and files tended to necessitate longer correction times. Previous research has explored CI issues in closed-source projects within industrial settings, and our study takes a more specialized approach by investigating the intricacies of compilation within embedded systems. Our research addresses a critical gap in understanding the CI process within this niche area.

\subsection{Dependency errors}
Software bugs have been studied extensively, yet rarely have studies been conducted on dependency bugs. Kerzazi, Khomh, and Adams~\cite{kerzazi2014automated} study build failures. Our previous work shows that dependency errors take a significant portion of industrial build failures~\cite{fu2022prevalence}. Fischer-Nielsen et al.~\cite{fischer2020forgotten} characterize the dependency bugs in the Robot Operating System (ROS) and study the pervasiveness and potential solutions of these bugs. Khazem et al.’s research~\cite{khazem2018making} into how portable software can be given changes to the toolchain (C/C++ compiler) or standard C library demonstrates the challenge of producing reproducible platform-independent software. Zakaria et al.~\cite{zakaria2022mapping} explore mechanisms to find dependencies of High-Performance Computing (HPC) in the context of a taxonomy of software distribution. Resolving dependency errors is a known subject in software configuration management~\cite{ratti2018conceptual}. To date, there has been a lack of sufficient work in identifying the precise locations of dependency errors. 

\subsection{Fault localization with log parsing}
Fault localization techniques are widely proposed and developed in a broad spectrum. Wong et al.'s study catalog provides a comprehensive overview of such techniques and discusses critical issues and concerns pertinent to software fault localization~\cite{wong2016survey}. 

Log parsing is crucial for fault localization, but manual log structuring using rule-based approaches is time-consuming and not scalable~\cite{rodrigues2021clp}. Academic research has proposed automated log analysis techniques like LogRAM, DeepLog, and LogTools~\cite{dai2020logram, du2017deeplog, zhu2019tools}. However, limited research has been conducted in complex industrial settings.

Logs serve as a valuable resource for diagnosing system failures. Prior research has explored reconstructing failed executions or differentiating execution flows based on log data~\cite{fu2014digging, lou2010mining, zhou2018fault}. Another line of work focuses on identifying root-cause-related log messages by comparing logs during failure periods with reference logs without failures. Notable approaches in this realm include LogCluster~\cite{lin2016log}, Log3C~\cite{he2018identifying}, and Onion~\cite{zhang2021onion}. The aforementioned studies lack systematic analysis within the industrial CI system, let alone a specific focus on compilation errors.

Ziftci and Reardon~\cite{ziftci2017broke} study integrating fault localization techniques into a continuous integration system at Google. Their work indicates that various fault localization techniques are unsuitable for rapid development cycles, as at Google. Hassan and Zhang~\cite{hassan2006using} carry out a study on a large software project at IBM, using classifiers to predict whether a build would pass a certification process.

Our study builds upon these prior efforts by combining log parsing and commit tracing techniques to gain insights into compilation errors within an industrial CI context.

\section{Study Design}\label{sec:studydesign}
\subsection{CI diagnostics solution}
\begin{figure*}
    \centering  
    \includegraphics[scale=.5]{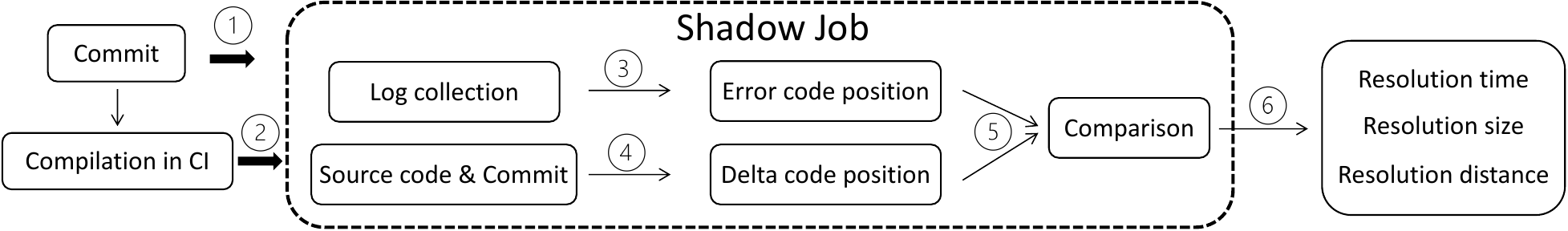}
    \caption{Shadow Job}
    \label{fig:shadowjob}
\end{figure*}

We propose a CI diagnostics solution that analyzes the outcomes of each CI invocation without disrupting the original pipeline. This solution, which analyzes the build outcomes, is referred to as a~\emph{shadow job}. A shadow job is a duplicate of a configuration that runs in parallel with the main job. Its design enables it to gather data from each execution step without disrupting the original CI pipeline. It is essential that our diagnostic solution does not disrupt the regular CI process; furthermore, it should be efficient. 

In highly active product environments, adding new steps to the main job can be both expensive and resource intensive. A more practical approach is to design a separate CI diagnostic solution that operates in parallel, utilizing a low-cost server, since the allocated hardware and software resources are already in use when the main job is triggered. This parallel design enables the implementation of the diagnostic solution without impacting the main job, ensuring scalability and the independent addition of new functionalities to the shadow job. By decoupling the diagnostic process from the main job, the solution can operate efficiently, providing valuable insights without disrupting the primary development workflow.

As illustrated in Fig.~\ref{fig:shadowjob}, the shadow job collects data from commits, source code, and compilation logs in the CI system. In step 1, the shadow job gathers commits that merge into the main branch of the product. Within each commit, the shadow job collects the patch that triggers the compilation errors and the patch that fixes the compilation errors for later comparison. In step 2, the shadow job collects the compilation build logs. Step 1 and step 2 are interconnected, since one commit with multiple changes in step 1 triggers the compilation build log in step 2. The shadow job collects and analyzes the build logs based on merged commits. If a commit is merged, the shadow job follows patches to locate each build log triggered by each patch.

In step 3, if a build fails, the shadow job extracts the error messages and the corresponding error code position (line number in the code) to obtain the error code position. In step 4, the shadow job extracts the error correction patch's delta code position (line number in the code). Similar to steps 1 and 2, steps 3 and 4 are interconnected.

In step 5, the shadow job compares the errors and delta code positions. In step 6, we calculate the resolution time, resolution size, and resolution distance.

We design the following studies to answer five research questions about the prevalence of different compilation errors (RQ1), the resolution time, size, and distance of different types of fixes (RQ2--4), and whether these metrics are correlated (RQ5): \\

\subsubsection{RQ1}\label{RQ1}
\textit{What are the most common compilation errors in an industrial CI system?}\\
As our previous research~\cite{fu2022prevalence} shows, compilation errors are a major contributor to CI build errors. We want to understand industrial CI systems' most common compilation errors and some of their causes. If the build fails, we collect the corresponding commit Change-ID for each build. In addition, the corresponding exception types are stored. 

Next, error messages from the build log are systematically categorized into distinct classes, encompassing Dependency, Syntax, Type Mismatch, Semantic, and Others~\cite{seo2014programmers}, facilitating a comprehensive analysis of their characteristics. Because builds with more than two compilation errors are rare, we can analyze data from patches that fix broken builds quite reliably, as most fixes target one or two compilation errors. Furthermore, we assume that the patch leading to the first successful build after a series of failures is the patch that fixes the compilation error.

\subsubsection{RQ2}\label{RQ2}
\textit{What is the resolution time for fixing different compilation errors?}\\
Here we measure developers' elapsed time on fixing different compilation error types, corresponding to the collected compilation error builds. Additionally, we highlight and describe example instances of fixes. 

It is important to note that resolution time can be influenced by various human factors and the specific error-handling processes employed by different companies or organizations. Within large and complex organizations, multiple developers may be involved in resolving errors, which can potentially introduce permission issues when code changes span across different functions.

As discussed in Section~\ref{sec:background}, there is often a mismatch between software and hardware design at the early stages of development. Consequently, even if software developers successfully compile and test their code locally with the simulator, there is still a possibility of compilation failures in the CI environment. In such cases, developers may need to wait for an updated version of the hardware design in the CI system.

Our research is centered on establishing the relationship between the commit that initiates the compilation error and the commit responsible for its resolution. In contrast, other studies, such as the one by Seo et al.~\cite{seo2014programmers}, focus on the time between the completion of the first failing build and the start of the subsequent successful build. This approach may be useful in analyzing the overall cost of resolving compilation errors and avoids underestimating the resolution time. Our method accounts for the fact that the fix process has to be managed and delegated, and we assume that intermediate commits may correspond to unrelated tasks. Therefore, we specifically consider the failing build preceding the successful build within a series of builds.

\begin{figure}[H]
    \centering  
    \includegraphics[scale=.26]{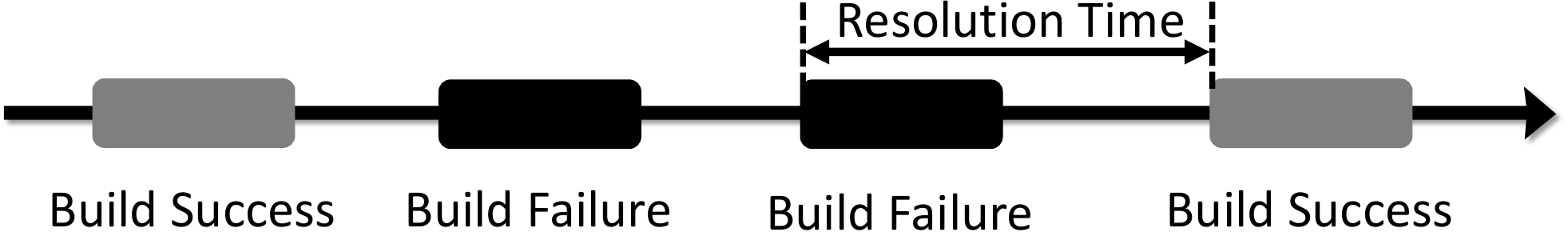}
    \caption{Resolution Time}
    \label{fig:resolutiontime}
\end{figure}

Figure~\ref{fig:resolutiontime} illustrates how we measure the resolution time. In this example, there are at least two failing builds before the first successful build. We measure the time from the beginning of the lastest failing build to the beginning of the successful build, capturing the duration of the resolution process. While this may under-approximate the time taken to resolve a problem, we believe it is more accurate in our situation, taking into account the embedded development process.

\subsubsection{RQ3}\label{RQ3}
\textit{What is the size of error corrections?}\\ \
In addition to analyzing resolution time, we also look at the resolution size of each fix. We utilize the information in the failed build log, including the commit ID and the faulty line of code in the faulty files. With this information, we can extract the faulty file.
Next, we proceed to extract the first subsequent successful compilation build. We can determine the resolution size by comparing the faulty file with the corresponding file from the successful build. The number of deletions and additions to the file determines the resolution size as follows: 

Size 0 indicates either a permission change or an alteration of a binary file. Size 1 indicates a single line of code being deleted or added. Size 2 indicates a single line of code being modified, one line being added and another one deleted, or two lines being deleted or added, respectively.

\subsubsection{RQ4}\label{RQ4}
\textit{What are the resolution distances for different error types?}\\
We aim to analyze the distance between the lines of code where the fixes are made and the lines of code indicated by the error messages. This information helps us identify the code sections that need to be modified in order to address the compilation errors effectively. Therefore, we investigate the feasibility of automatically locating error corrections. The failed build log includes a commit id and faulty line of code of faulty files. With this information, we extract faulty lines of code and their build environment to reproduce the faulty build. We also extract the patch within one merged commit that leads to the failed CI build caused by compilation error as the error code base with the number $E$. We then extract the next patch that leads to the CI build passing the compilation stage as the fixing code base, with the number $F_n$. Finally, we compare the error code base and fixing code base to have the code difference. In addition, we also extract the indication of the error line of code from the CI failure log. 

We calculate the resolution distance as shown in Equation~\ref{eq:resolutionDistance}. We may discover multiple fixing locations in one patch with line numbers $F_1$, $F_2$, ..., and $F_n$. We then take the minimum of all distances between the compilation error location $E$ and the nearest fix $F_i$. Thus, the resolution distance $D$ is the minimum value in a tuple $(D_1, D_2, ..., D_n)$. 
\begin{equation}\label{eq:resolutionDistance}
\begin{aligned}
    D &= \min(D_1, D_2, \dots, D_n), \quad \text{where} \\
    D_i &= |F_i - E| \quad \text{for } i = 1, 2, \dots, n
\end{aligned}
\end{equation}

\subsubsection{RQ5}\label{RQ5}
\textit{Which types of compilation errors are suitable to apply fault localization and automatic program repair in industrial embedded systems?}\\
We want to understand the correlation between resolution distance and resolution size. Based on their frequency of occurrence, we focus our investigation on the top four error types. These error types are deemed significant for potential automated fault localization and automated program repair. Furthermore, we have gathered sufficient data to explore any potential correlation between these error types and their resolution attributes.

To calculate the correlation matrix based on resolution distance and each error type's corresponding resolution size and time, we use the Pearson correlation coefficient~\cite{cohen2013applied} as a statistical measure that quantifies the strength of the linear relationship between different continuous variables.

\subsection{Data collection}
Our data collection was conducted on a single project at Ericsson over the course of a year. This project is highly active and involves the development of embedded software.

As our data were collected from the centralized CI system, our data collection lacks errors encountered when a developer compiles locally. We therefore do not count errors that developers can resolve locally before committing.

\section{Study Results}\label{sec:study-results}
\subsection{Results for RQ1}
To answer RQ1, we automatically collect the build logs through the shadow job in step 2 in Fig.~\ref{fig:shadowjob}. We map the error messages into 14 compilation error types based on the Yocto project~\cite{khandelwal2017enhancement} compiler configuration files. The Yocto Project is an open-source collaboration project that provides templates, tools, and methods to help create custom Linux-based systems for embedded products. 

\begin{mdframed}[style=insight, frametitle={Key insight of RQ1:}]
    Dependency issues contribute to 76\,\% of all compilation errors, highlighting the challenges in reconciling the disparities between the embedded CI system and local development environments.
    \end{mdframed}

\begin{figure}
    \centering  
    \includegraphics[scale=.54]{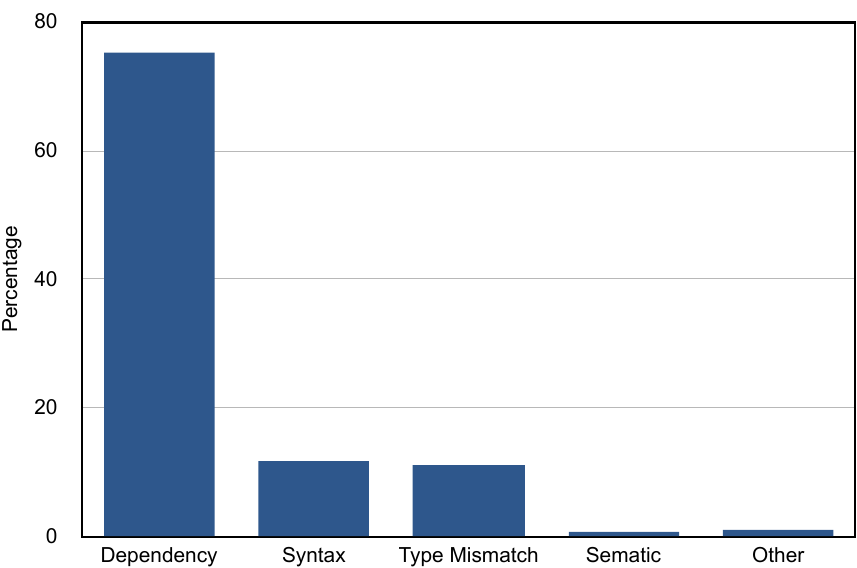}
    \caption{Compilation error class}
    \label{fig:errorclass}
\end{figure}

Table~\ref{tab:CompilationError} shows 14 types of compilation errors and their proportion distribution. The ratio of compilation error types ranges from 0.36\,\% to 40.05\,\%. Subsequently, we classify the different error types into 5 classes, as illustrated in Fig.~\ref{fig:errorclass}.\\

\begin{table}
    \captionsetup{justification=centering} 
    \caption{Compilation Error Statistics}
    \centering
    \renewcommand{\arraystretch}{1.2} 
    \setlength{\tabcolsep}{7.6pt} 
    \begin{tabular}{@{}cllr@{}}
    \toprule
    No. & Error type                  & Class         & {\%} \\
    \midrule
    1   & was not declared            & Dependency    & 40.05 \\
    2   & has no member named         & Dependency    & 20.18 \\
    3   & expected X before Y token   & Syntax        & 11.77 \\
    4   & does not name a type        & Dependency    & 8.89  \\
    5   & no declaration matches      & Type mismatch & 8.36  \\
    6   & no such file or directory   & Dependency    & 2.76  \\
    7   & ld returned                 & Dependency    & 2.21  \\
    8   & invalid conversion          & Type mismatch & 1.53  \\
    9   & unused variable             & Dependency    & 1.14  \\
    10  & does not have any field named & Type mismatch & 0.82  \\
    11  & cannot allocate an object of & Semantic      & 0.73  \\
    12  & of non-class type           & Other         & 0.71  \\
    13  & cannot convert              & Type mismatch & 0.49  \\
    14  & static assertion failed     & Syntax        & 0.36  \\
    \bottomrule
    \end{tabular}
    \label{tab:CompilationError}
\end{table}

As can be seen from Table~\ref{tab:CompilationError}, the top five types of compilation errors are responsible for 89.25\,\% of all errors. Error type \textsl{was not declared} takes the majority of compilation errors, with 40.05\,\%. This is significantly higher than other pure software projects reported by Seo et al.~\cite{seo2014programmers}. This error message indicates that the compiler has encountered a reference to a variable or function that has not been declared. The reason this cannot be found in the local environment before commit is that there is a gap between the local environment and the CI environment, especially at the early stage of development. 

We want to understand the implications of the gap between the local and CI development environment. We adopt the classification of compilation errors from the literature~\cite{seo2014programmers} to categorize different error types into classes—the classes of each error type are shown as the second column in Table~\ref{tab:CompilationError}. Grouping these subclasses of faults allows us to identify common characteristics and similarities, indicating that they can be addressed using similar automated solutions. \textit{Dependency} refers to errors that are missing packages, libraries, or other resources. Error types like \textsl{was not declared}, \textsl{has no member named}, \textsl{does not name a type}, \textsl{no such file or directory}, \textsl{ld returned}, and \textsl{unused variable} fall into class \textit{Dependency.} \textit{Syntax} refers to errors that occur when the compiler or interpreter encounters code that does not follow the rules or syntax of the programming language. Error types like \textsl{expected X before Y token} and \textsl{static assertion failed} fall into class \textit{Syntax}. \textit{Type mismatch} refers to a mismatch between the types of variables or expressions used in an operation or assignment. Therefore, error types like \textsl{no declaration matches}, \textsl{invalid conversion}, \textsl{does not have any field named}, and \textsl{cannot convert} fall into class \textit{Type mismatch}. \textit{Semantic} error indicates a problem with the meaning or interpretation of the code rather than its syntax or dependencies. Therefore, \textsl{cannot allocate an object of} falls into class \textsl{Semantic}. Error type \textsl{of non-class type} indicates that the variable involved in the operation is not a class or struct type. We categorize it as \textit{Other}.

Fig.~\ref{fig:errorclass} shows the distribution of different classes. It shows the \textit{Dependency} class takes 76\,\%. Our finding further substantiates our previous research~\cite{fu2022prevalence}. The proportion is also significantly higher than purely software projects reported by Seo et al.~\cite{seo2014programmers}. 
The gap between local and CI development environments with embedded systems in the loop is much larger then we expect. In addition, we also see that \textit{Type mismatch} and \textit{Syntax} have similar proportions, around 12\,\%. Our result is therefore different from Seo et al.~\cite{seo2014programmers} reporting that \textit{Type mismatch} is the second biggest class. 

\subsection{Results for RQ2}
\begin{figure}
    \centering
    \includegraphics[scale=.46]{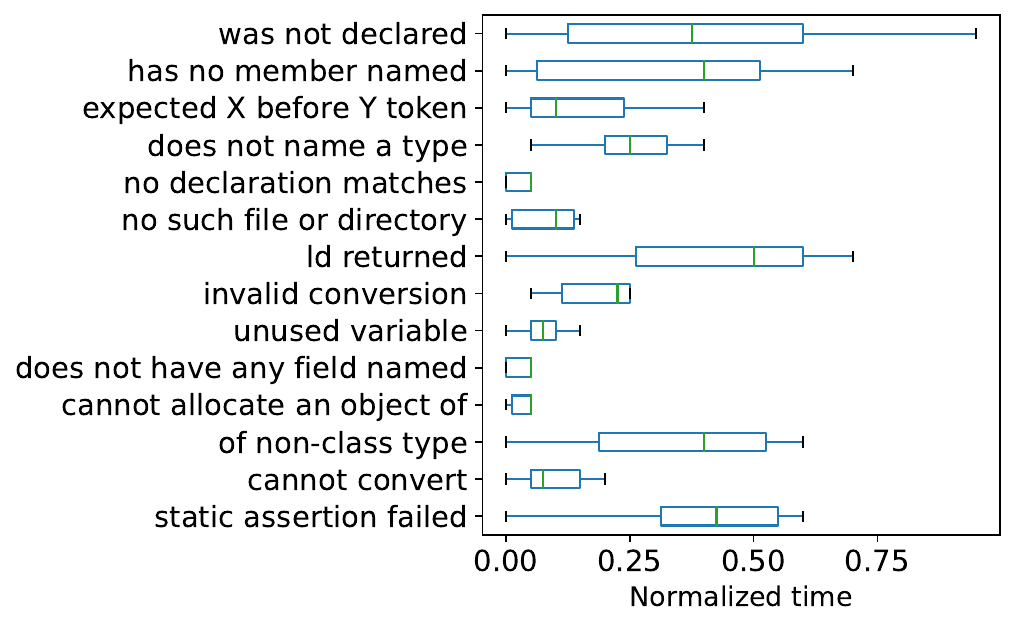}
    \caption{Resolution time statistics}
    \label{fig:timestatistics}
\end{figure}

This section examines each error type's resolution time. Fig.~\ref{fig:timestatistics} shows the box plot for resolution times. The box is bounded by the 25 and 75 percentiles, and the line within the box is the median value. We normalize resolution times to a scale of [0,1] for confidentiality reasons.

Errors that occur more frequently tend to require a longer resolution time. For example, the error types of \textsl{was not declared} and \textsl{has no member named} exhibit a noticeably longer resolution time compared to other error types. Additionally, their distribution has a higher variance than that of other error types. In our CI system, each modification involves various patches; these patches not only fix compilation errors but also introduce different functionalities. Consequently, a larger dataset leads to a wider distribution of resolution times. This aligns with the fact that the \textsl{was not declared} error accounts for 40.05\,\% of errors, as illustrated in Table~\ref{tab:CompilationError}. \\

\begin{mdframed}[style=insight, frametitle={Key insight of RQ2:}]
The observation that compilation errors require a significant effort to resolve despite their high frequency is counterintuitive and surprising. One would typically expect that the more frequently a specific type of error occurs, the faster the resolution process would be.
\end{mdframed}

Based on their frequency of occurrence, we focus our investigation on the top four error types, which collectively constitute 81\,\% of our dataset.\footnote{Due to the sensitive and proprietary nature of the industrial context, we are unable to share specific detailed data, including the total number of errors.} These error types are deemed significant for potential automated fault localization and automated program repair, given their substantial representation in our data.

Certain error types, such as \textsl{no declaration matches}, \textsl{no such file or directory}, and \textsl{unused variable}, are relatively simple to resolve, as indicated by shorter resolution times. As the frequency of errors decreases, it typically leads to a reduction in resolution time. Error types like \textsl{ld returned} (a linker error) take longer to fix due to their complexity. Linker errors often arise when a project has complex dependencies between different modules or libraries. These dependencies can be difficult to identify and resolve, especially in larger projects. In an embedded system, each hardware has its own specific set of libraries. Therefore, during local development, the software is typically compiled with the libraries specific to the hardware being used. This approach is more practical and efficient than providing cross-compilation for all different hardware configurations, which would be costly and time-consuming.

\subsection{Results for RQ3}
\begin{figure}
    \centering  
    \includegraphics[scale=.48]{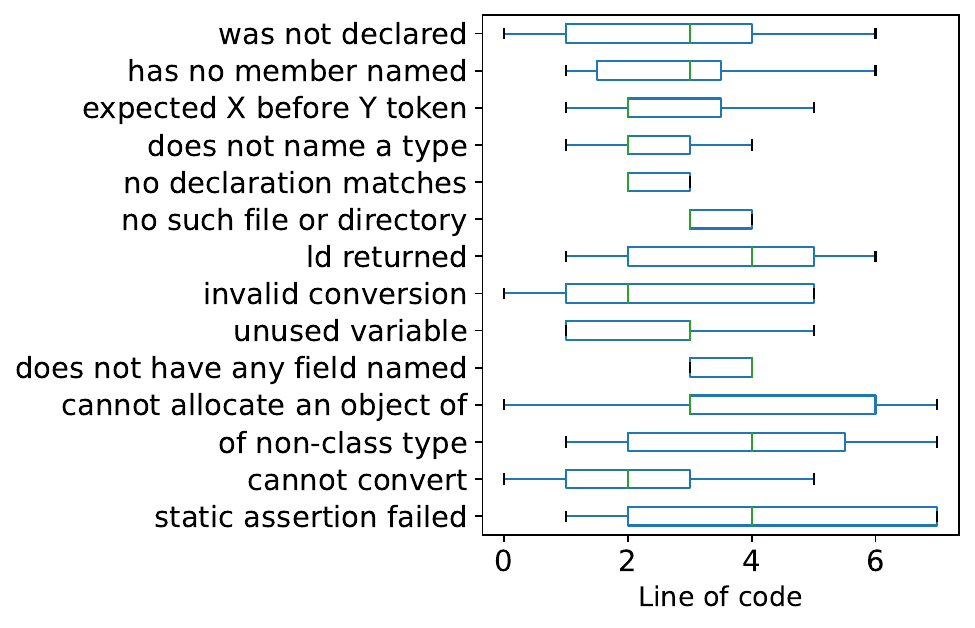}
    \caption{Resolution size statistics}
    \label{fig:sizestatistics}
\end{figure}

\begin{figure}
    \centering  
    \includegraphics[scale=.54]{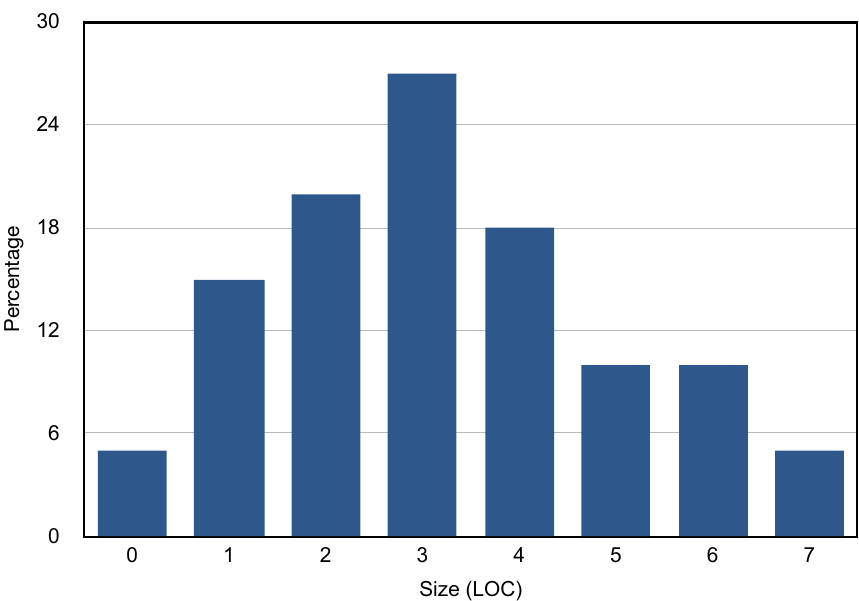}
    \caption{Resolution size distribution}
    \label{fig:sizedis}
\end{figure}

In this section, we examine the resolution size of each error type. Remember that we count both additions and deletions here, so four lines typically correspond to two deleted and two added lines (which can be an edit within two lines); changes in binary files count as zero lines. Similar to Fig.~\ref{fig:timestatistics}, Fig.~\ref{fig:sizestatistics} displays a box plot illustrating the distribution of resolution sizes for each error type.\footnote{In case two quartiles have the same number of data points, the quartile boundaries are calculated using the average between the largest element of the lower quartile and the smallest element of the higher quartile, which sometimes results in fractions.}
Additionally, Fig.~\ref{fig:sizedis} presents the distribution of resolution sizes for all errors, categorized into 0--7 lines of code.\\

\begin{mdframed}[style=insight,frametitle={Key insight of RQ3:}]
    The majority of resolution size to compilation errors are changes between 1 and 4 lines.
\end{mdframed}

As shown in Fig.~\ref{fig:sizestatistics}, the most prevalent error type, identified as \textsl{was not declared}, tends to have a relatively small resolution size, typically ranging from 1 to 4 lines of code. The top 5 error types generally have a smaller resolution size, usually fewer than 4 lines of code. The error type \textsl{ld returned} leads to a noticeably larger resolution size and time, which is due to the complexity of the linking process. Linking processes can be intricate, particularly in a large embedded system with numerous dependencies. Additionally, linker errors may arise because of platform-specific differences. Reproducing the same platform can be challenging when the developer does not have the same environment as the CI system.

As depicted in Fig.~\ref{fig:sizedis}, the most frequent resolution size observed is 3 (in approximately 27\,\% of the cases). The second most common resolution size is 2, constituting around 20\,\% of the occurrences. Larger resolution sizes such as 5, 6, and 7 are much less common.

\subsection{Results for RQ4}

\begin{figure}
    \centering  
    \includegraphics[scale=.48]{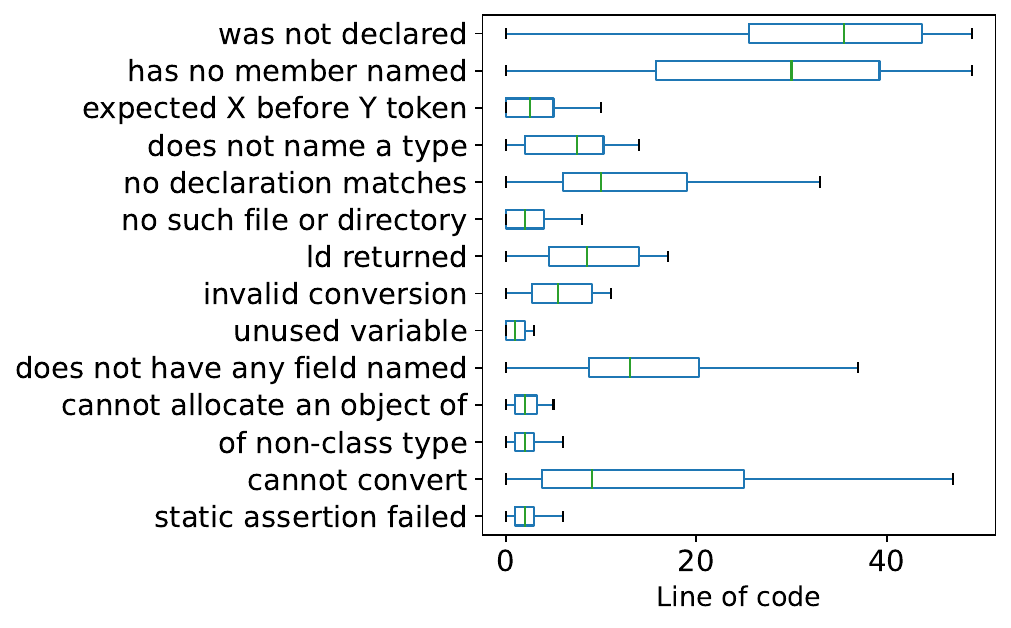}
    \caption{Resolution distance statistics}
    \label{fig:distancestatistics}
\end{figure}

We measure the resolution distance in RQ4 in Section~\ref{RQ4}. Fig.~\ref{fig:distancestatistics} displays a box plot illustrating the distribution of resolution distance for each error type.

As shown in Fig.~\ref{fig:distancestatistics}, the error types \textsl{was not declared} and \textsl{has no member named} exhibit a relatively large distance between them. The \textsl{was not declared} error type demonstrates a larger resolution distance. The difference in resolution distance between the \textsl{was not declared} and \textsl{has no member named} errors can be attributed to the specific nature of the corrections required. In the case of the \textsl{was not declared} error, the correction typically involves adding a declaration at the beginning of the file, rather than modifying the specific code line indicated by the error message.

The \textsl{was not declared} error type typically occurs 25 to 42 lines away from the error message location, while the \textsl{has no member named} error type primarily appears 15 to 38 lines away from the error message location. This discrepancy can be attributed to the fact that header files often provide declarations for functions and classes that are defined in separate source files. As a result, resolving the \textsl{was not declared} error may require modifications at the beginning of the file, leading to a larger resolution distance.\\

\begin{mdframed}[style=insight,frametitle={Key insight of RQ4:}]
    Many frequently occurring error types are close to the reported location and therefore good targets for automatic fault localization. However, the top two most frequent error types require a full syntactic source code analysis to locate automatically due to their large resolution distance. 
\end{mdframed}

\subsection{Results for RQ5}
\begin{figure*}[ht!]
    \begin{adjustwidth}{-1in}{-1in}  
    \centering
    \begin{tabular}[c]{cccc}
      \begin{subfigure}[b]{0.475\columnwidth}
        \centering
        \begin{adjustbox}{valign=M}
        \includegraphics[width=\textwidth]{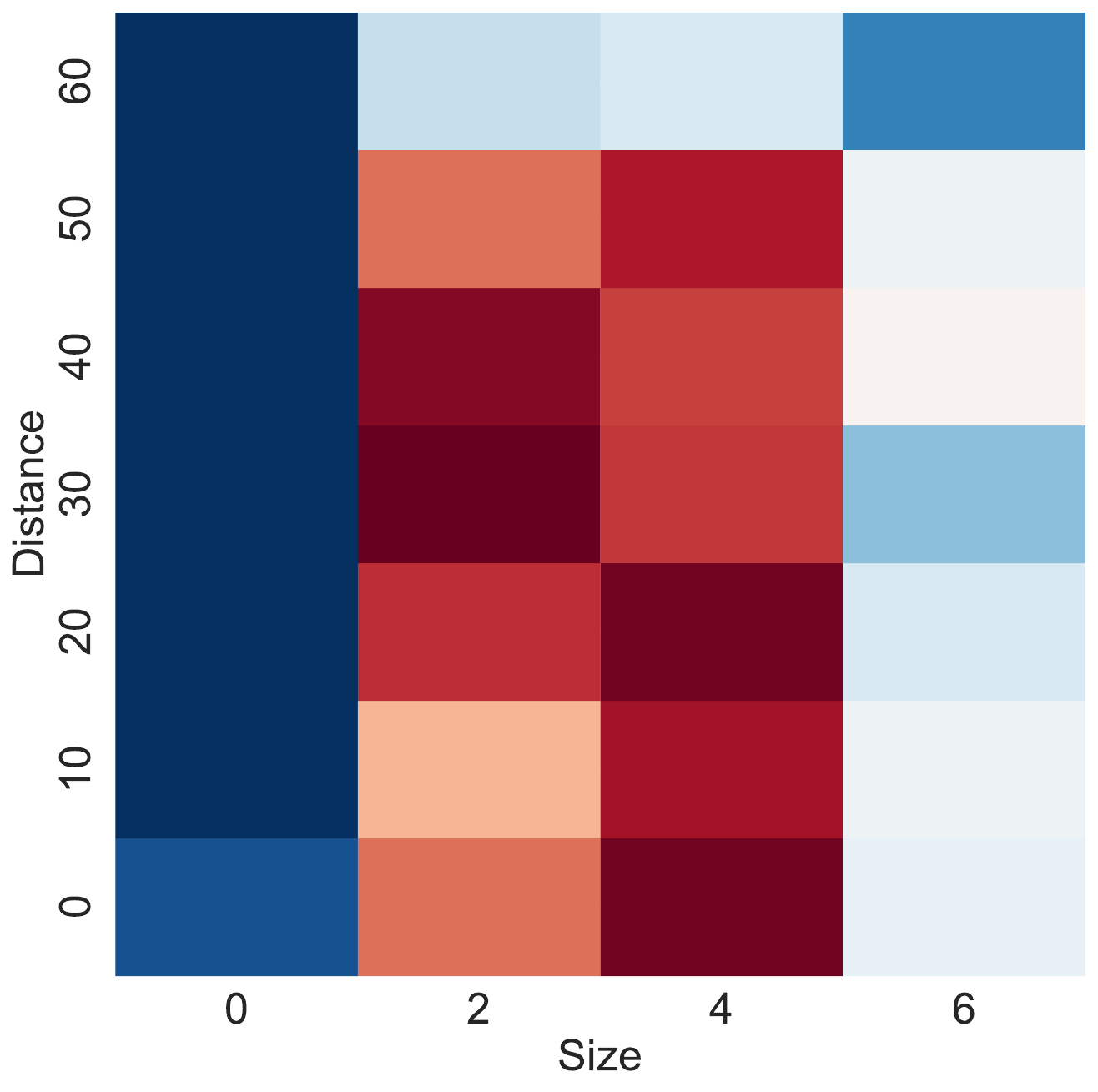}
        \end{adjustbox}
        \caption{was not declared}  
      \end{subfigure}&
      \begin{subfigure}[b]{0.475\columnwidth}
        \centering
        \begin{adjustbox}{valign=M}
        \includegraphics[width=\textwidth]{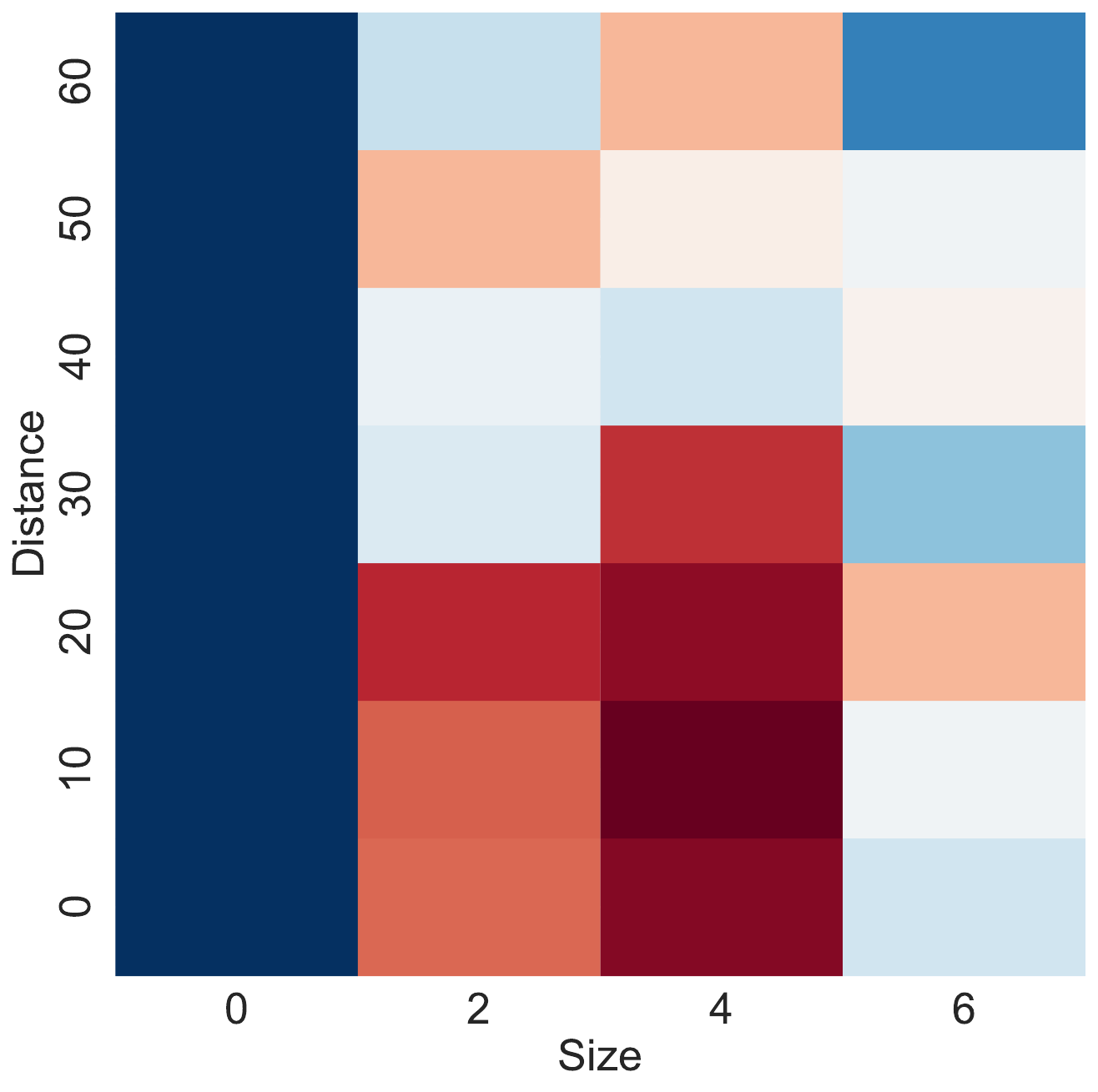}
        \end{adjustbox}
        \caption{has no member named} 
      \end{subfigure}&
      \begin{subfigure}[b]{0.475\columnwidth}
        \centering
        \begin{adjustbox}{valign=M}
        \includegraphics[width=\textwidth]{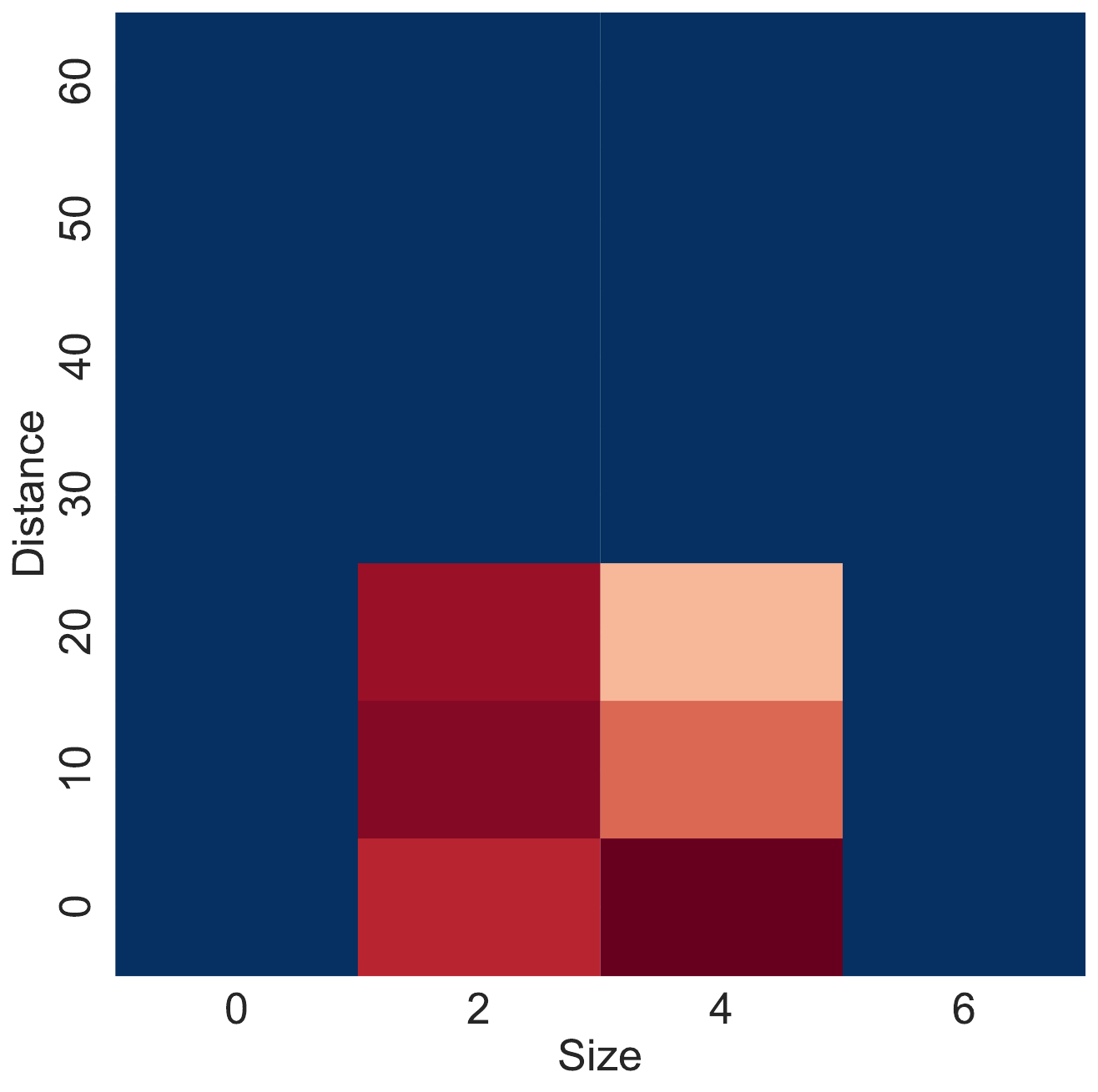}
        \end{adjustbox}
        \caption{expected X before Y token}
      \end{subfigure}&
      \begin{subfigure}[b]{0.525\columnwidth}
        \centering
        \begin{adjustbox}{valign=M}
        \includegraphics[width=\textwidth]{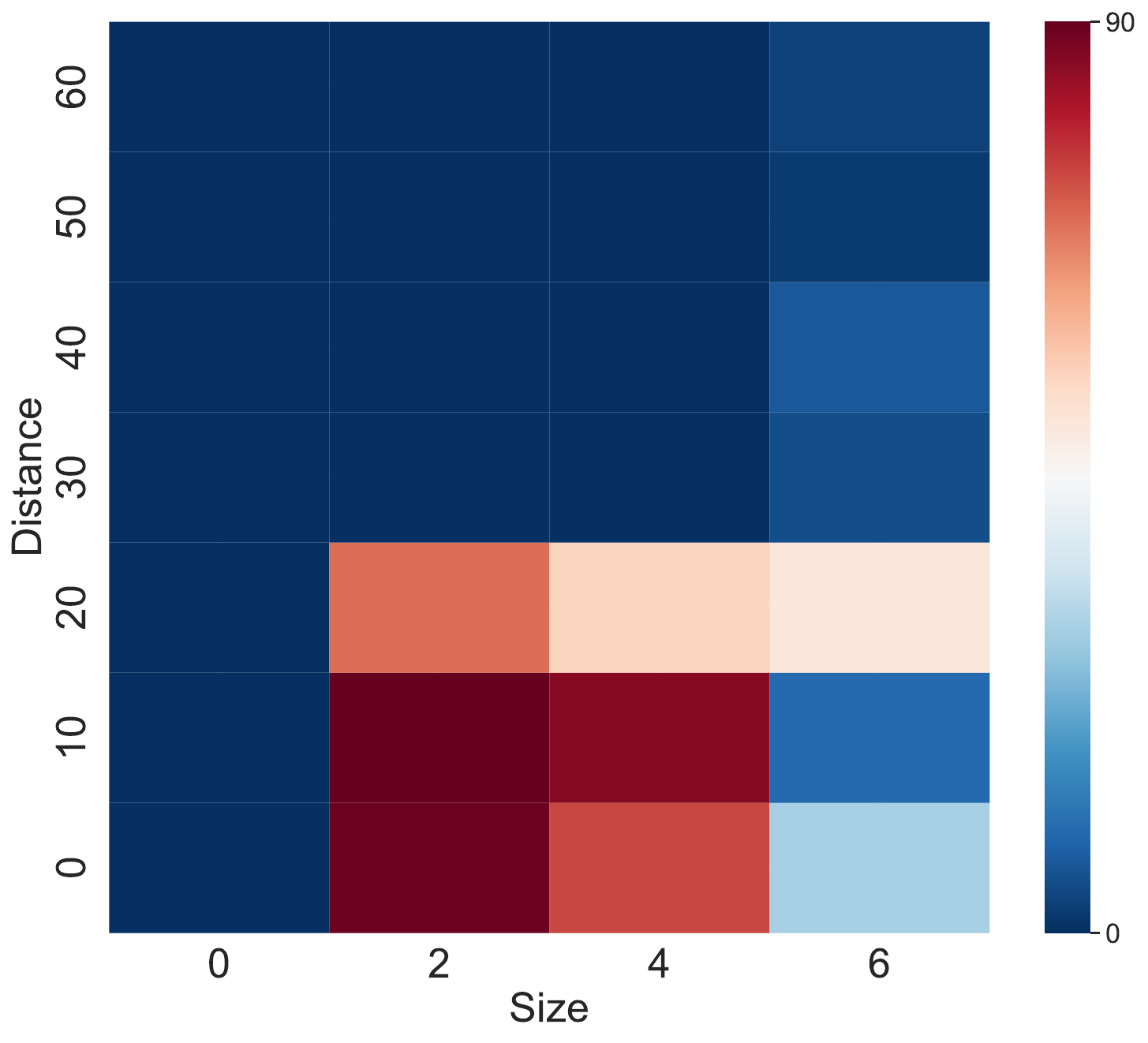}
        \end{adjustbox}
        \caption{does not name a type}
      \end{subfigure}
    \end{tabular}
    \end{adjustwidth}  
    \caption{Heat map showing the distribution of fix size and distance for the four most common compilation errors}
    \label{fig:heatmap1-4}
  \end{figure*}

The results of RQ2 in Section~\ref{RQ2} and RQ3 in Section~\ref{RQ3} indicate that the top 4 error types are more representative and exhibit greater consistency. Moreover, the top 4 error types account for 80.89\,\% of the dataset, making it reasonable to focus our discussion primarily on these error types.

The top 4 error types demonstrate a relatively smaller resolution size and a distance of usually up to 50 lines from the location of the compilation error. While they do not require extensive modification to be fixed, many fixes for the top 2 error types are ``far away'' from the error message location in terms of resolution distance and size. This finding suggests that the error types \textsl{was not declared} and \textsl{has no member named} may be challenging for automatic fault localization techniques.

Due to their larger resolution distances and sizes, these error types may require more manual intervention, code analysis, and extensive code modifications to rectify. As a result, automated fault localization methods that primarily rely on localizing errors based on error messages or limited code regions may not be effective in accurately pinpointing the root cause of these particular error types.

\begin{table}[htbp]
    \centering
    \caption{Correlation Between Key Attributes}
    \setlength{\tabcolsep}{1.1pt}
    \renewcommand{\arraystretch}{1.5}
\resizebox{\columnwidth}{!}{
    \begin{tabular}{@{}l@{}rrr@{}}
        \toprule
        \textbf{Error} & \textbf{Distance--Size} & \textbf{Distance--Time} & \textbf{Size--Time} \\
        \midrule
        was not declared    & $0.227$ & $0.034$ & $-0.287$ \\
        has no member named & $0.047$ & $-0.091$ & $0.155$ \\ 
        expected X before Y token & $0.053$ & $0.114$ & $0.057$ \\
        does not name a type & $0.220$ & $-0.040$ & $0.082$ \\
        \midrule
        All errors & $-0.069$ & $0.192$ & $0.077$ \\
        \bottomrule
    \end{tabular}
}
    \label{tab:correlation}
\end{table}
We evaluate the correlation between resolution distance, size, and time for each of the four most common error types. Table~\ref{tab:correlation} presents the correlation values for the top 4 error types and overall error types, measuring 3 types of correlations: distance to size, distance to time, and size to time. Fig.~\ref{fig:heatmap1-4} illustrates the heatmap for the top 4 error types based on resolution sizes and distances. 

In Table~\ref{tab:correlation}, the first row represents the correlation values for the error type~\textsl{was not declared}. The highest positive correlation is between distance and size, with a value of $0.227$. The highest negative correlation is between size and time, with a value of $-0.287$. These values are all low, showing that resolution distance, size, and time are almost unrelated to each other. A small fix that is close to the location of the compilation error may still be hard to find for a human.

The findings align with the heatmap presented in Fig.~\ref{fig:heatmap1-4} (a)-(d), where each cell's shade corresponds to the color bar on the right side of the heatmap. The bottom shade represents the minimum value, while the top shade represents the maximum value.

In Fig.~\ref{fig:heatmap1-4} (a), the first column displays a nearly consistent color, indicating that errors rarely occur when the resolution size is 0. Moving to the second column, where the resolution size is two lines of code, the color changes based on the frequency of occurrence. Specifically, when the resolution size is two lines of code, the majority of resolution distances cluster around 30 lines of code.

\begin{mdframed}[style=insight,frametitle={Key insight of RQ5:}]
    Resolution distance, size, and time are independent of each other. 
\end{mdframed}

The fixes for the error type~\textsl{has no member named} tend to cluster around a resolution size of 2--4 lines of code and a resolution distance of 0--20 lines of code, as shown in Fig.~\ref{fig:heatmap1-4} (b). Compared to the error type~\textsl{was not declared}, fixing~\textsl{has no member named} appears to require relatively less effort. The heatmaps in Fig.~\ref{fig:heatmap1-4} (c) and (d) indicate slightly varying resolution distances for different errors, but the resolution sizes for the most common errors are quite similar.

\subsection{Threats to validity}
\subsubsection{Internal}
In a complex industrial embedded CI system, dependency-related failures are prevalent~\cite{fu2022prevalence}. Extracting data from real-use code prevents us from rerunning executions with specific environments after discarding outdated dependency build packages, limiting our adoption of fault localization and automated program repair in such a CI system.

Furthermore, a single commit often includes multiple changes. When calculating resolution distance and size, we consider only the files indicated by the error message, potentially leading to occasional miscalculations. Given that most errors are related to dependency declarations, we find this approach suitable.

While our method of calculating resolution time aligns with the development process according to our judgment, a more precise measurement would involve accounting for working hours, capturing the actual time designers spent fixing errors.

The current approach to computing change size, encompassing both code and binary changes, may pose challenges. To enhance clarity and ensure the visibility of binary changes, it is suggested to consider a separation between changes in code and binaries.

\subsubsection{External}
In calculating both resolution time and resolution distance, it is inevitable that the human factor influences the study results. Variances in development pace, work methodologies, developers' habits, and the distribution of teams and developers are key factors that can affect the outcomes. To mitigate these influences, conducting in-depth analyses through interviews with developers emerges as a valuable approach.

In addition to this, the correlation coefficient choice depends on data characteristics and assumed relationships between variables. Opting for the Pearson correlation, suited to linear relationships in our data, may limit the study scope to linear dependencies. Exploring alternative metrics like min-max resolution time and distance, employing machine learning models, and utilizing different analyses such as SHAP~\cite{lundberg2017unified} can provide a broader perspective on metric dependencies.

\section{Discussion}\label{sec:disucssion}
Our analysis of over 40000 builds reveals that a significant portion of industrial build failures can be attributed to a dependency on its hardware-in-the-loop CI system. The integration of software with hardware prototypes in CI poses a challenge in bridging the gap between the CI and local development environments. The complexities introduced by hardware-in-the-loop testing and the difficulty in reproducing the hardware-in-the-loop environment accurately within the CI system contribute to build failures that are hard to resolve. Addressing this mismatch is crucial for improving the stability and reliability of the build process in an embedded industrial CI system.

Our finding of resolution time reveals that defects that occur more frequently tend to require a longer resolution time. Therefore, it would be highly advantageous to prioritize the development of automated solutions for resolving the most frequently occurring error types. 

A lengthy resolution time for the top 4 errors does not necessarily mean a large resolution size. In fact, the majority of fixes involve changes of only 1 to 4 lines of code. This quantification of resolution size inspires developers by highlighting the success of small fixes. Existing automated program repair approaches designed for one-line changes can be applicable in addressing these compilation errors efficiently.

Further examination of the top 4 error types shows that resolution time, size, and distance are independent attributes. 
This suggests that compilation errors possess inherent statistical characteristics that render them more conducive to automated detection and remediation.

That says no specific type of compilation error lends itself to a single simple strategy for automated fixes. Due to this, we should prioritize frequently occurring errors. In light of the absence of such data, it is advantageous to prioritize techniques that specifically target the most common error types.

In summary, our study suggests that automated fault localization and program repair efforts show promise for the identified error types. However, addressing resolution distance and size separately is crucial for developing effective techniques in these contexts. Our results indicate a significant potential for implementing automatic fault localization techniques for compilation errors, emphasizing the need for tailored approaches.

\section{Conclusion}\label{sec:conclusion}
We conducted a statistical analysis on over 40000 builds in a highly active product using a scalable CI diagnostics solution called ``Shadow Job.'' We extracted compilation errors into 14 error types and classified them into 5 classes. The results highlight the significance of the compilation step in industrial-embedded development CI systems. 76\,\% of compilation errors are due to dependencies between hardware and software and a mismatch between the embedded environment (offered by the CI system) and the local development environment.

Our study also categorized, quantified, and analyzed compilation errors based on three key attributes: resolution time, size, and distance. More frequent compilation errors require more time to be resolved, although the size of the fixes is always small, mostly within one or two lines of changes. The fix location can be relatively far away from the location of the compilation error, but resolution time, size, and distance are not correlated among each other. 

The fact that the five most frequent compilation errors make up 76\,\% of all compilation errors suggests that a collection of specialized strategies to prevent or automatically locate and repair these five errors can be useful.

\section{Future work}\label{sec:futurework}
In future work, a key focus will be on the practice of automatic fault localization and automatic program repair, specifically targeting the reduction of compilation errors caused by configuration issues and dependency problems. Large modular systems often have numerous dependencies reflected in their configuration settings, which are typically stored in different formats that have evolved independently.

To further enhance our "Shadow Job" implementation, we aim to expand its capabilities beyond diagnosis and providing valuable error location information. Our goal is to develop an automated solution that can automatically fix identified errors. For instance, for multi-line compilation errors in the C language, one approach could involve using a neural network-based approach, such as DeepFix~\cite{gupta2017deepfix}. Alternatively, a novel approach would involve leveraging a large language model like Keep~\cite{xia2023keep} to address these errors.

By incorporating advanced techniques such as neural networks or large language models, we can enhance the automated resolution capabilities of our system, facilitating the efficient and effective handling of multi-line code fixes. This approach holds promise for reducing the manual effort required to fix complex compilation errors and improving the overall development process.

We also intend to extend the study to other projects and/or companies. This will address the external validity threats and apply the findings to more CI pipelines to validate our results.

\bibliographystyle{ieeetr}
\balance
\bibliography{references.tex}
\end{document}